\documentclass[%
 reprint,
 twocolumn,
superscriptaddress,
 amsmath,amssymb,
 aps,
prx,
longbibliography,
]{revtex4-2}

\usepackage{xcolor}
\usepackage{graphicx}% Include figure files
\usepackage{dcolumn}% Align table columns on decimal point
\usepackage{bm}% bold math
\begin{document}

\title{Jamming of Bidisperse Frictional Spheres}

\author{Ishan Srivastava}
 \email{isriva@lbl.gov}
\affiliation{Center for Computational Sciences and Engineering, Lawrence Berkeley National Laboratory, Berkeley, California 94720, USA}
\affiliation{Sandia National Laboratories, Albuquerque, New Mexico 87185, USA}

\author{Scott A. Roberts}
\affiliation{Sandia National Laboratories, Albuquerque, New Mexico 87185, USA}

\author{Joel T. Clemmer}
\affiliation{Sandia National Laboratories, Albuquerque, New Mexico 87185, USA}

\author{Leonardo E. Silbert}
\affiliation{School of Math, Science, and Engineering, Central New Mexico Community College, \\
Albuquerque, New Mexico 87106, USA}%

\author{Jeremy B. Lechman}
\affiliation{Sandia National Laboratories, Albuquerque, New Mexico 87185, USA}

\author{Gary S. Grest}
\affiliation{Sandia National Laboratories, Albuquerque, New Mexico 87185, USA}

%\date{\today}
             %  but any date may be explicitly specified% It is always \today, today,

\begin{abstract}
By generalizing a geometric argument for frictionless spheres, a model is proposed for the jamming density $\phi_J$ of mechanically stable packings of bidisperse, frictional spheres. The monodisperse, $\mu_s$-dependent jamming density $\phi_J^{\mathrm{mono}}(\mu_s)$ is the only input required in the model, where $\mu_s$ is the coefficient of friction. The predictions of the model are validated by robust estimates of $\phi_J$ obtained from computer simulations of up to $10^7$ particles for a wide range of $\mu_s$, and size ratios up to 40:1. Although $\phi_J$ varies nonmonotonically with the volume fraction of small spheres $f^s$ for all $\mu_s$, its maximum value $\phi_{J,\mathrm{max}}$ at an optimal $f^{s}_{\mathrm{max}}$ are both $\mu_s$-dependent. The optimal $f^{s}_{\mathrm{max}}$ is characterized by a sharp transition in the fraction of small rattler particles.
\end{abstract}

\maketitle
%\Srivastava{Please write comments using your last name commands in the TeX file directly---see this comment as an example}
Granular materials jam at a range of densities $\phi_J$ depending on particle shape~\cite{torquato2010,smith2014,salerno2018}, size dispersity~\cite{brouwers2006,farr2009,hopkins2013,desmond2014,koeze2016,prasad2017a,pillitteri2019,petit2020a,hara2020} and interparticle interactions~\cite{silbert2010,santos2020}. In the simplest case of monodisperse spheres, $\phi_J^{\mathrm{mono}}\!\simeq\!0.64$ is the random closed packed density for a mechanically stable packing of frictionless particles~\cite{ohern2003,torquato2010}. However, particle surface roughness introduces friction upon their contact and jamming can occur at a density as low as $\phi_J^{\mathrm{mono}}\!\simeq\!0.55$~\cite{jerkins2008,baule2018,santos2020}. Conversely, bidispersity in particle sizes can increase $\phi_J$ to a theoretical maximum $\phi_J\!\simeq\!0.87$ for frictionless spheres~\cite{kansal2002,brouwers2006,farr2009,hopkins2013,prasad2017a,petit2020a,hara2020}. However, only limited studies~\cite{goncu2013} have explored the combined role of friction and dispersity on the jamming of spheres. A key impediment is the difficulty of simulating large-scale, mechanically stable, jammed packings of frictional particles in the limit of marginal rigidity \cite{blumenfeld2005} using previously established methods that render such packings prone to instabilities at low confining pressures near $\phi_J$~\cite{silbert2010}. Additional challenges include inefficient neighbor finding algorithms for large particle size ratio systems~\cite{intveld2008,ogarko2012}.

Dispersity in particle sizes results in a rich structural diversity in particulate materials such as granular materials~\cite{danisch2010,cantor2018,mutabaruka2019,petit2020a}, colloids~\cite{hermes2010,voigtmann2011}, emulsions~\cite{ricouvier2017} and geophysical materials~\cite{jerolmack2019a}. Maximizing $\phi_J$ in such materials through size dispersity is important in various applications such as battery electrodes~\cite{srivastava2020a}, cement~\cite{masoero2012} and chocolate~\cite{blanco2019}. The mechanics of such materials strongly correlates with $\phi_J$~\cite{kumar2016a,liu2019b,hara2020}, and size dispersity provides a powerful knob to optimize their mechanical properties. Furthermore, $\phi_J$ also critically governs the equilibrium~\cite{spangenberg2014,pednekar2018,cantor2018} and nonequilibrium~\cite{guy2020} rheology of dense, bidisperse particulate materials. Therefore, beyond their fundamental jamming characteristics, a robust estimation of $\phi_J$ for bidisperse particulate materials with realistic interparticle interactions is important towards advancing our knowledge of their mechanics and rheology.

Here, we use pressure-controlled simulations to simulate mechanically stable jammed packings of frictional, bidisperse spheres for a wide range of particle size ratios $2\!\le\!\alpha\!\le\!40$ and volume fraction of small particles $0\!\leq\!f^s\!<\!1$. Here, $\alpha \!=\! d_l/d_s$ and $f^s\!=\!N_sd_{s}^{3}/(N_sd_{s}^{3}+N_ld_{l}^{3})$, where $d_s$ and $d_l$ are the diameters, and $N_s$ and $N_l$ are the numbers of small and large spheres, respectively. The effect of $\mu_s$ on $\phi_J$ is demonstrated through $\mu_s$-dependent state diagrams on $\left(\phi_J,f^s\right)$ axes by varying the coefficient of friction over several orders of magnitude $0\!\leq\!\mu_s\!\leq\!0.5$, where $\mu_s\!=\!0$ corresponds to frictionless particles. Similar to previous findings~\cite{prasad2017a,petit2020a,hara2020}, we find that $\phi_J$ varies nonmonotonically with $f^s$ and attains a maximum $\phi_{J,\mathrm{max}}$ at $f^{s}_{\mathrm{max}}$. At large $\alpha$, this variation exhibits a sharp transition at $f^{s}_{\mathrm{max}}$. Although $\phi_J$ varies similarly with $f^s$ for all $\mu_s$, the optimal $\phi_{J,\mathrm{max}}$ and $f^{s}_{\mathrm{max}}$ depend significantly on $\mu_s$. The optimal $\phi_{J,\mathrm{max}}$ decreases systematically with $\mu_s$, while requiring a larger $f^{s}_{\mathrm{max}}$ to attain its optimal value. Through these findings, we propose a generalized Furnas model~\cite{furnas1929}---originally proposed for notionally placing small particles in the available volume left by large particles without reference to mechanical constraints or particle properties other than infinite size ratio---that accurately predicts $\mu_s$-dependent $\phi_J$ for large $\alpha$, although small deviations from the model are observed for frictional packings at $f^{s}<f^{s}_{\mathrm{max}}$ and very large $\alpha$. Remarkably, the generalized model only requires $\mu_s$-dependent monodisperse jamming density $\phi_J^{\mathrm{mono}}(\mu_s)$ as an input to accurately predict $\phi_J$; thus $\phi_J^{\mathrm{mono}}(\mu_s)$ appears to encode the mechanical stability constraint of the packing in such a way that the available volume for small frictional particles within the space left over by large frictional particles is correspondingly adjusted. Lastly, we find that for $f^s\!<\!f^s_{\mathrm{max}}$ all the small particles rattle in the void space of the jammed network of large particles, and the fraction of such rattlers drops sharply at $f^{s}_{\mathrm{max}}$. Unlike frictionless particles, the rattler fraction does not drop to zero for frictional particles.

Three key features distinguish the present simulations from similar previous works: (i) we use pressure-controlled simulations that guarantee mechanical stability, unlike commonly used volume controlled methods~\cite{dagois-bohy2012,smith2014}; (ii) we create packings up to an unprecedentedly high $\alpha\!=\!40$ in order to probe the large size ratio limit; as a result, we have created the densest known bidisperse jammed packings for frictionless and frictional spheres; (iii) we provide statistically robust estimates of $\phi_J$ by fixing a large value of $N_l\!=\!1000$ in all simulations. This constraint resulted in a large number of small particles up to $N_s\!\simeq\!3\!\times\!10^{7}$, which were simulated through large-scale discrete element method using the software LAMMPS~\cite{plimpton1995}.

A typical jamming simulation starts with a dilute non-overlapping collection of spheres of material density $\rho\!=\!1$ at $\phi\!=\!0.05$ containing $N_l$ large particles and $N_s$ small particles determined from the chosen values of $\alpha$ and $f^s$. The initial simulation cell is periodic and cubic; however, its triclinic nature allows shear distortions in addition to volume deformation. At $t\!=\!0$ a pressure $p_a$ is applied such that the total stress on the system is $\boldsymbol{\sigma}_a\!=\!p_a\boldsymbol{I}$, where $\boldsymbol{I}$ is the identity matrix. Particles interact with damped Hookean and the tangential frictional forces. The tangential spring stiffness $k_n$ is set equal to the normal spring stiffness $k_t$, and a Coulomb coefficient of interparticle friction $\mu_s$ sets the sliding frictional force~\cite{srivastava2021,srivastava2019}. The damping at contacts is set by normal and tangential velocity damping coefficients $\gamma_{n,t}=0.5$. Time is normalized by the characteristic timescale of collision $t_{c}\!=\!\pi(2k_{n}/m_{s}\!-\!\gamma_{n}^{2}/4)^{-1/2}$, where $m_s$ is the mass of the small particle. The simulation time step is set to $0.02t_c$. The stress in this setup is scaled by $k_n/d_s$ and energy is scaled by $k_nd_s^{2}$.

\begin{figure}[t]
\includegraphics[width=\columnwidth]{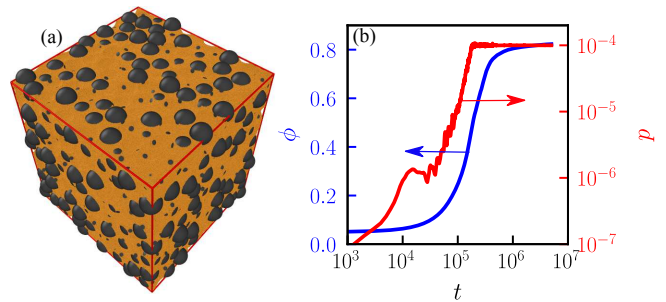}% Here is how to import EPS art
\caption{(a) Snapshot of a jammed packing containing bidisperse particles with $\alpha\!=\!40$, $f^s\!=\!0.32$ and $\mu_s\!=\!0.3$. This system contain $N_l\!=\!10^{3}$ large particles (grey) and $N_s\!\simeq\!2.6\!\times\!10^{7}$ small particles (gold). (b) The transient evolution of $\phi$ and $p$ as a function of simulation time $t$.}
\label{fig1}
\end{figure}

Under the action of $p_a$, the system steadily compacts and its internal pressure $p$ steadily increases towards $p_a$, as shown in Fig.~\ref{fig1}(b). The applied pressure is fixed at a low value $p_a\!=\!10^{-4}$ in all simulations to model jamming in the asymptotic hard-particle regime~\cite{roux2000}. The inertial equations of motion for the particles and the simulation cell are described in detail in Ref.~\cite{srivastava2021}. After transient evolution, the system eventually achieves a jamming density $\phi_J$ when $p$ balances $p_a$, as shown in Fig.~\ref{fig1}(b). We set a criterion to terminate a simulation when the average kinetic energy per particle falls below $10^{-11}$ or if the simulation has proceeded for at least $5\!\times\!10^{6}$ and up to $10^{7}$ steps, which were found sufficient to achieve a stable jammed packing within a reasonable computational time. Beyond pressure, the pressure-controlled simulation method also ensures that the deviatoric internal stress is equal to its externally applied value upon jamming, i.e., zero~\cite{smith2014,santos2020}. This is achieved by allowing shear distortions of the simulation cell~\cite{santos2020,srivastava2021}. As a result, mechanically stable jammed packings are created at a specified pressure $p_a$, similar to soft particle jamming using the variable-cell structural optimization method~\cite{smith2014,srivastava2017}, and strictly-jammed hard particle packings using the adaptive shrinking cell method~\cite{hopkins2013}.

Traditional neighbor finding and interprocessor communication algorithms in particle-based simulations are inefficient for modeling systems with large size dispersity~\cite{intveld2008,ogarko2012}. Here, we adapt and implement an algorithm originally developed for colloidal mixtures~\cite{intveld2008} to simulate bidisperse granular systems for large size ratios up to $\alpha\!=\!20$. We also include results from jamming simulations at very large $\alpha\!=\!40$ by adapting a recently developed neighbor finding algorithm that efficiently simulates granular systems of large size ratios~\footnote{We used a recently proposed algorithm to speedup particle contact detection~\cite{shire2020}. Separate spatial binning of small and large particles reduces the time to find potential neighbors, while a hierarchical search from small to large particles reduces the interprocessor communication.}. An example of a $\alpha\!=\!40$ packing that would have been computationally prohibitive to simulate using traditional neighbor finding methods is shown in Fig.~\ref{fig1}(a).

About a century ago, Furnas predicted a nonmonotonic relationship $\phi_J(f^s)$ in the limit $\alpha \!\to\! \infty$ based on a simple geometric model~\cite{furnas1929}. At low $f^s$ up to a critical $f^{s}_{\mathrm{max}}$, $\phi_J\!=\!\phi_J^{\mathrm{mono}}/(1\!-\!f^s)$ is an increasing function of $f^s$ resulting from \emph{unjammed} small particles occupying the void space in the network of large particles jammed at $\phi_J^{\mathrm{mono}}$, which is the jamming density for monodisperse frictionless spheres. At $f^{s}_{\mathrm{max}}$, small particles are also jammed at $\phi_J^{\mathrm{mono}}$, and the system achieves it highest packing density $\phi_{J,\mathrm{max}}$. For $f^s\!>\!f^s_{\mathrm{max}}$, the small particles form a percolating jammed network in which large particles are suspended (and jammed), and the $\phi_J$ decreases monotonically with $f^s$ as $\phi_J\!=\!\phi_J^{\mathrm{mono}}/[f^s\!+\!\phi_J^{\mathrm{mono}}(1\!-\!f^s)]$. For the jamming density $\phi_J^{\mathrm{mono}}\!\simeq\!0.64$, the model predicts $f^s_{\mathrm{max}}\!\simeq\!0.26$ and $\phi_{J,\mathrm{max}}\!\simeq\!0.87$.

\begin{figure}[t]
\includegraphics[width=\columnwidth]{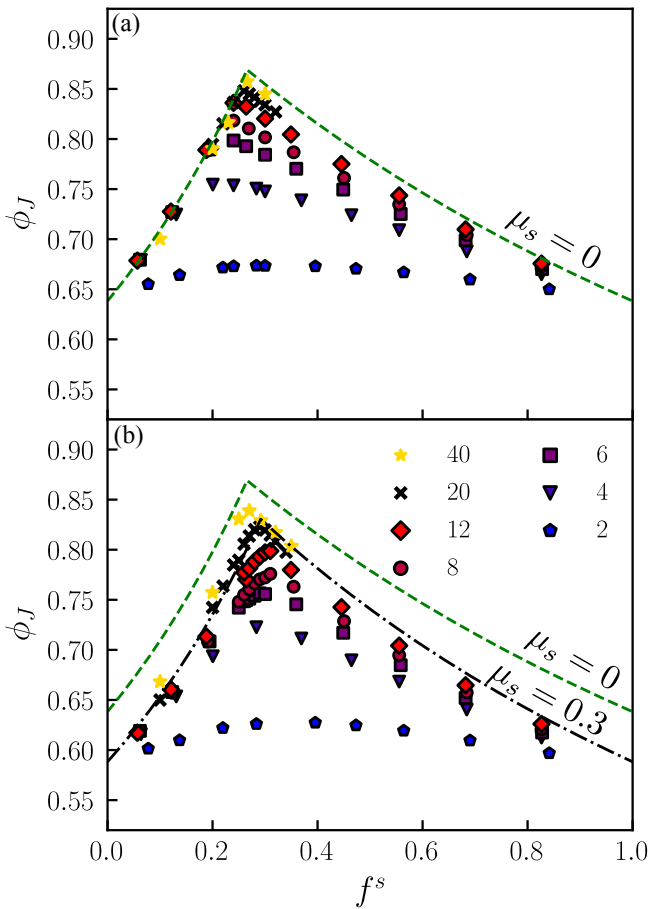}% Here is how to import EPS art
\caption{Variation of $\phi_J$ with $f^s$ for (a) frictionless particles and (b) particles with friction $\mu_s\!=\!0.3$ at various $\alpha$ (see legend in (b)). The green dashed and the black dot-dashed lines represent the model predictions for $\phi_{J}$ for $\mu_s\!=\!0$ and $\mu_s\!=\!0.3$ respectively.}
\label{fig2}
\end{figure}

The predictions of this model have been tested in simulations of bidisperse, frictionless spheres up to $\alpha\!=\!12$~\cite{prasad2017a,petit2020a}. Using large scale simulations, we created jammed, bidisperse, frictionless packings up to $\alpha\!=\!40$. Figure~\ref{fig2}(a) shows the variation of $\phi_J$ with $f^s$ for various $\alpha$ along with the predictions from the Furnas model. At low $\alpha$, $\phi_J$ varies continuously and nonmonotonically with $f^s$, and this variation exhibits a sharp transition at high $\alpha$, similar to previous observations~\cite{prasad2017a,petit2020a}. For $\alpha\!=\!20$, $\phi_J$ follows the predictions of the Furnas model and closely approaches $\phi_{J,\mathrm{max}}\!=\!0.869$ at $f^s_{\mathrm{max}}\!=\!0.265$, which were estimated by the Furnas model using $\phi_J^{\mathrm{mono}}\!\simeq\!0.638$ obtained at $f^s\!=\!0$. We also simulated a packing for $\alpha\!=\!40$ containing $N_s\!\simeq\!2.3\!\times\!10^{7}$ particles at the purported optimal $f^s\!=\!0.265$, as shown in Fig.~\ref{fig2}(a). The value $\phi_J\!=\!0.857$ obtained in this large simulation is, to our knowledge, the densest bidisperse jammed packing created using computer simulations.

A similar variation of $\phi_J$ with $f^s$ is observed for frictional particles, as shown in Fig.~\ref{fig2}(b) for $\mu_s\!=\!0.3$. Similar to frictionless particles, $\phi_J$ varies continuously at low $\alpha$, whereas the variation exhibits a sharp transition at higher $\alpha$. However, $\phi_J$ at all $\alpha$ and $f^s$ is substantially lower than $\phi_J$ for frictionless particles. For monodisperse spheres ($f^s\!=\!0$), we obtained $\phi_J^{\mathrm{mono}}\!\simeq\!0.588$. In addition to lower $\phi_J$ compared to frictionless particles, the maxima in $\phi_J$ occurs at a higher $f^s$. Following the data for $\alpha\!=\!20$ in Fig.~\ref{fig2}(b), we find that the maximum $\phi_J\!=\!0.823$ occurs at $f^s\!=\!0.29$, which is higher than $f^s$ required to maximize $\phi_J$ for frictionless particles.

Such differences in $\phi_J(f^s)$ based on $\mu_s$ are explained by generalizing the Furnas model to include $\mu_s$-dependence of $\phi_J^{\mathrm{mono}}$. At low $f^s$, large particles form a jammed network, but at a reduced $\phi_J^{\mathrm{mono}}(\mu_s)$~\cite{silbert2010}. Upon increasing $f^s$, small particles fill the voids of the jammed network, but only up to the same reduced $\phi_J^{\mathrm{mono}}(\mu_s)$, at which point $\phi_{J,\mathrm{max}}$ is obtained at $f^{s}_{\mathrm{max}}$. Using $\phi_J^{\mathrm{mono}}=0.588$ for $\mu_s=0.3$ in the generalized Furnas model, $f^s_{\mathrm{max}}=0.292$ and $\phi_{J,\mathrm{max}}=0.83$ are predicted, which are remarkably close to the values obtained from simulations shown in Fig.~\ref{fig2}(b) for $\alpha=20$. A large jamming simulation for $\alpha=40$ containing $N_s\simeq2.6\times10^{7}$ particles for $f^s=0.292$ results in $\phi_J=0.83$, as shown in Fig.~\ref{fig2}(b), which confirms these predictions.

For the largest simulated frictional packings at $\alpha\!=\!40$, $\phi_J$ is slightly greater than the model prediction for $f^{s}<f^{s}_{\mathrm{max}}$, and the deviation from the model increases with $f^{s}$, as shown in Fig.~\ref{fig2}(b). A similar but reduced deviation from the model is also observed for $\alpha\!=\!20$ packings. This is caused by the large particles packing at a density greater than $\phi_J^{\mathrm{mono}}(\mu_s)$, but still bounded by the frictionless $\phi_J^{\mathrm{mono}}$. This effect vanishes for $f^{s}>f^{s}_{\mathrm{max}}$ when small particles also participate in jamming, and $\phi_J$ agrees well with the model. As monodisperse particles can jam at a range of densities depending on the protocol~\cite{luding2016,bertrand2016,baule2018}, we expect that in the region near $f^{s}_{\mathrm{max}}$, the same would be true for bidisperse packings.

\begin{figure}[t]
\includegraphics[width=\columnwidth]{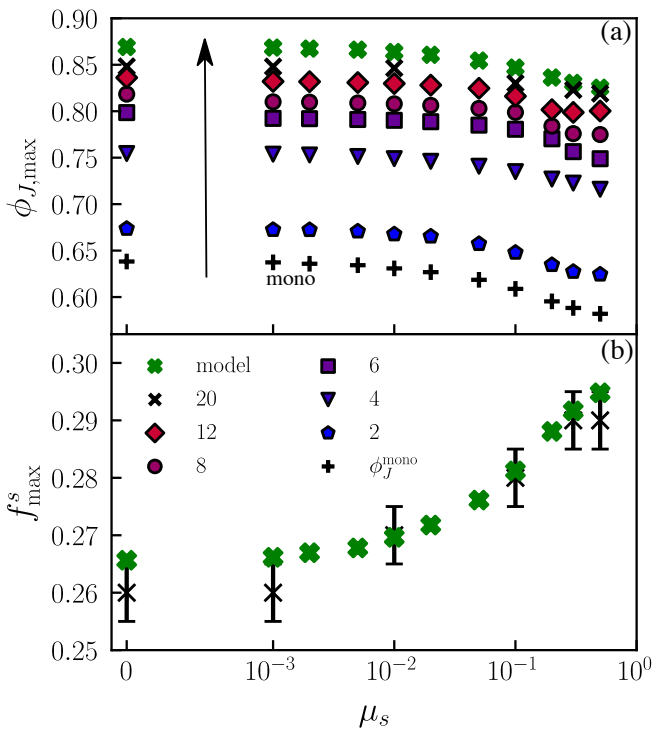}
\caption{Variation of (a) $\phi_{J,\mathrm{max}}$ and (b) $f^{s}_{\mathrm{max}}$ with $\mu_s$ for various $\alpha$ (see legend in (b)). The arrow shows the direction of increasing $\alpha$. The green filled `$\boldsymbol{\mathrm{x}}$' in (a) and (b) denote model predictions, and `$\bm{+}$' in (a) denote $\phi_J^{\mathrm{mono}}$.}
\label{fig3}
\end{figure}

To test the predictions of the generalized Furnas model for $\phi_{J,\mathrm{max}}$ as a function of $\mu_s$, we first obtain monodisperse $\phi_J^{\mathrm{mono}}(\mu_s)$ shown in Fig.~\ref{fig3}(a). These values are consistent with previous studies~\cite{silbert2010,srivastava2019,santos2020}. Upon substituting $\phi_J^{\mathrm{mono}}$ in the generalized Furnas model, theoretical estimates for $\phi_{J,\mathrm{max}}$ are obtained, as depicted in Fig.~\ref{fig3}(a). The theoretical predictions are tested against $\phi_{J,\mathrm{max}}$ obtained from simulations for various $\alpha$, as shown in Fig.~\ref{fig3}(a). At lowest $\alpha\!=\!2$, the calculated $\phi_{J,\mathrm{max}}$ are much lower than the theoretical maxima, but still substantially higher than $\phi_J^{\mathrm{mono}}$. As $\alpha$ increases, the calculated $\phi_{J,\mathrm{max}}$ steadily increases towards its theoretical maxima and is within $2\%$ of its maximum value for $\alpha=20$. For high friction, $\alpha=20$ appears to be sufficient to nearly attain the theoretical maxima $\phi_{J,\mathrm{max}}$. Jamming at low friction requires much longer simulations to completely remove the kinetic energy of the particles. The presence of friction between particles provides additional constraints to the particle motion~\cite{santos2020}, thus enabling a quicker approach to jamming.

Similar to $\phi_{J,\mathrm{max}}$, the theoretical predictions for $\mu_s$-dependent $f^{s}_{\mathrm{max}}$ compare well with simulations for $\alpha=20$, as shown in Fig.~\ref{fig3}(b). As $\mu_s$ increases, a larger fraction of small particles is required to achieve the highest packing density, with the highest friction case requiring $3\%$ higher volume of small particles than the frictionless case.

\begin{figure}[t!]
\includegraphics[width=\columnwidth]{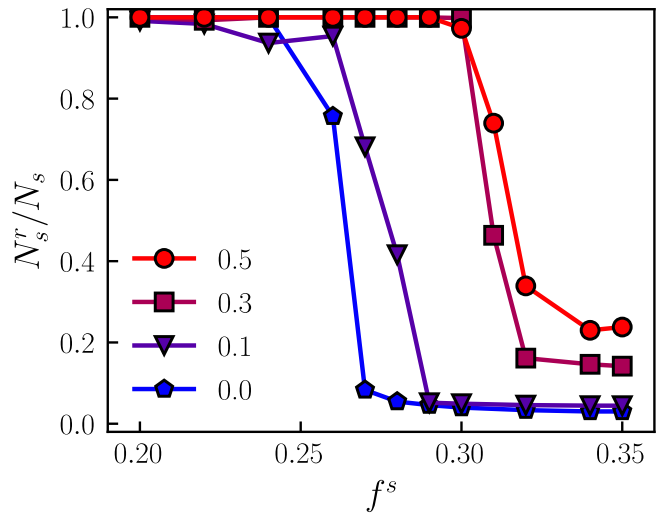}% Here is how to import EPS art
\caption{Variation of the fraction of small rattler particles $N_{s}^{r}/N_s$ with $f^s$ for four $\mu_s$ (see legend) and $\alpha=20$.}
\label{fig4}
\end{figure}

Previous studies on jammed bidisperse frictionless particles have demonstrated a sharp transition in the structural properties across $f^{s}_{\mathrm{max}}$~\cite{prasad2017a,petit2020a}. Upon increasing $f^s$ at high $\alpha$, the fraction of small particles that are rattlers drops rapidly at $f^{s}_{\mathrm{max}}$~\cite{prasad2017a}. We analyze the fraction of small rattler particles to highlight a similar structural transition for frictional bidisperse packings. A particle is a rattler if it has too few contacts to contribute to the mechanical stability of the packing, i.e., the particle is locally unjammed~\cite{donev2004}. We follow a recursive method~\cite{donev2004} to identify if a particle $i$ is a rattler based on the criteria: $Z_i<3$ for frictionless particles and $Z_i<2$ for frictional particles, which emerges from constraint counting arguments~\cite{santos2020}. In Fig.~\ref{fig4}, the fraction of small rattler particles $N_{s}^{r}/N_s$ is plotted as a function of $f^s$ for four values of $\mu_s$, where $N_{s}^{r}$ is the total number of small rattler particles. Below a critical $\mu_s$-dependent $f^s$ that corresponds well with $f^{s}_{\mathrm{max}}$ shown in Fig.~\ref{fig3}(b), almost all of the small particles are rattlers within a jammed network of large particles. As $f^s$ is increased beyond $f^{s}_{\mathrm{max}}$, the fraction of small rattlers rapidly drops, and the majority of small particles participate in the mechanical backbone of the packing. Nearly all small particles are locally jammed for frictionless packings, whereas up to $22\%$ small particles are rattlers at high friction. These findings are consistent with previous studies on rattlers in monodisperse frictional packings near the jamming point~\cite{shundyak2007,santos2020}. The fraction of rattlers at high friction is similar to its value for monodisperse packings~\cite{santos2020}, thus indicating that small particles form a percolating jammed network with suspended (and locally jammed) large particles.

We have demonstrated that $\phi_J$ for bidisperse frictional particles is well-predicted by a simple $\mu_s$-dependent model for large size ratios. The model compares well with simulations that were used to create large mechanically stable bidisperse frictional packings at unprecedentedly high $\phi_J$. For frictional packings at $f^{s}<f^{s}_{\mathrm{max}}$ and $\alpha\!\gtrsim\!20$, $\phi_J$ is slightly larger than the model prediction, possibly due to the lubricating effect of the small particles. Friction is found to reduce $\phi_J$ for all $f^s$. At large $\alpha$, $\phi_J$ varies nonmonotonically with $f^s$ and exhibits a sharp transition at $f^{s}_{\mathrm{max}}$, where its value is maximized. The $f^{s}_{\mathrm{max}}$ that maximizes $\phi_J$ increases with $\mu_s$, with frictional contacts requiring up to $3\%$ more small particles to achieve $\phi_{J,\mathrm{max}}$. The sharp transition in $\phi_J$ is also observed structurally, where the fraction of small rattler particles rapidly drops across $f^{s}_{\mathrm{max}}$.

The results presented here open avenues to extend our understanding of the physics of jamming---particularly the structure and mechanical stability of frictional packings~\cite{henkes2016,liu2021}---to bidisperse frictional packings at large size ratios. Our ongoing work on including additional modes of friction, such as rolling and twisting, support an extended generalization of the Furnas model that takes into account $\phi_J^{\mathrm{mono}}$ in the presence of these additional frictional modes~\cite{santos2020}. This suggests that a century-old model may be a valuable predictor of $\phi_J$ in mechanically-stable bidisperse particulate packings spanning a wide variety of interparticle interactions. Lastly, a recent study has discovered an additional jamming transition at low $f^s$ for large $\alpha$ in bidisperse frictionless particles~\cite{petit2020a}. Although such a transition was not observed in our pressure-controlled jamming simulations, we can not preclude its presence without a careful analysis of the path dependence of jamming in our systems.

Helpful discussions with Andrew Santos and Dan Bolintineanu are acknowledged. I.~S. acknowledges support from the U.S. Department of Energy (DOE), Office of Science, Office of Advanced Scientific Computing Research, Applied Mathematics Program under contract No.~DE-AC02-05CH11231. This work was performed at the Center for Integrated Nanotechnologies, a U.S. DOE and Office of Basic Energy Sciences user facility. Sandia National Laboratories is a multimission laboratory managed and operated by National Technology and Engineering Solutions of Sandia, LLC, a wholly owned subsidiary of Honeywell International, Inc., for the U.S. DOE’s National Nuclear Security Administration under Contract No. DE-NA-0003525. This paper describes objective technical results and analysis. Any subjective views or opinions that might be expressed in the paper do not necessarily represent the views of the U.S. DOE or the United States Government.


\begin{thebibliography}{52}%
\makeatletter
\providecommand \@ifxundefined [1]{%
 \@ifx{#1\undefined}
}%
\providecommand \@ifnum [1]{%
 \ifnum #1\expandafter \@firstoftwo
 \else \expandafter \@secondoftwo
 \fi
}%
\providecommand \@ifx [1]{%
 \ifx #1\expandafter \@firstoftwo
 \else \expandafter \@secondoftwo
 \fi
}%
\providecommand \natexlab [1]{#1}%
\providecommand \enquote  [1]{``#1''}%
\providecommand \bibnamefont  [1]{#1}%
\providecommand \bibfnamefont [1]{#1}%
\providecommand \citenamefont [1]{#1}%
\providecommand \href@noop [0]{\@secondoftwo}%
\providecommand \href [0]{\begingroup \@sanitize@url \@href}%
\providecommand \@href[1]{\@@startlink{#1}\@@href}%
\providecommand \@@href[1]{\endgroup#1\@@endlink}%
\providecommand \@sanitize@url [0]{\catcode `\\12\catcode `\$12\catcode
  `\&12\catcode `\#12\catcode `\^12\catcode `\_12\catcode `\%12\relax}%
\providecommand \@@startlink[1]{}%
\providecommand \@@endlink[0]{}%
\providecommand \url  [0]{\begingroup\@sanitize@url \@url }%
\providecommand \@url [1]{\endgroup\@href {#1}{\urlprefix }}%
\providecommand \urlprefix  [0]{URL }%
\providecommand \Eprint [0]{\href }%
\providecommand \doibase [0]{https://doi.org/}%
\providecommand \selectlanguage [0]{\@gobble}%
\providecommand \bibinfo  [0]{\@secondoftwo}%
\providecommand \bibfield  [0]{\@secondoftwo}%
\providecommand \translation [1]{[#1]}%
\providecommand \BibitemOpen [0]{}%
\providecommand \bibitemStop [0]{}%
\providecommand \bibitemNoStop [0]{.\EOS\space}%
\providecommand \EOS [0]{\spacefactor3000\relax}%
\providecommand \BibitemShut  [1]{\csname bibitem#1\endcsname}%
\let\auto@bib@innerbib\@empty
%</preamble>
\bibitem [{\citenamefont {Torquato}\ and\ \citenamefont
  {Stillinger}(2010)}]{torquato2010}%
  \BibitemOpen
  \bibfield  {author} {\bibinfo {author} {\bibfnamefont {S.}~\bibnamefont
  {Torquato}}\ and\ \bibinfo {author} {\bibfnamefont {F.~H.}\ \bibnamefont
  {Stillinger}},\ }\bibfield  {title} {\bibinfo {title} {Jammed hard-particle
  packings: From kepler to bernal and beyond},\ }\href@noop {} {\bibfield
  {journal} {\bibinfo  {journal} {Rev. Mod. Phys.}\ }\textbf {\bibinfo {volume}
  {82}},\ \bibinfo {pages} {2633} (\bibinfo {year} {2010})}\BibitemShut
  {NoStop}%
\bibitem [{\citenamefont {Smith}\ \emph {et~al.}(2014)\citenamefont {Smith},
  \citenamefont {Srivastava}, \citenamefont {Fisher},\ and\ \citenamefont
  {Alam}}]{smith2014}%
  \BibitemOpen
  \bibfield  {author} {\bibinfo {author} {\bibfnamefont {K.~C.}\ \bibnamefont
  {Smith}}, \bibinfo {author} {\bibfnamefont {I.}~\bibnamefont {Srivastava}},
  \bibinfo {author} {\bibfnamefont {T.~S.}\ \bibnamefont {Fisher}},\ and\
  \bibinfo {author} {\bibfnamefont {M.}~\bibnamefont {Alam}},\ }\bibfield
  {title} {\bibinfo {title} {Variable-cell method for stress-controlled jamming
  of athermal, frictionless grains.},\ }\href@noop {} {\bibfield  {journal}
  {\bibinfo  {journal} {Phys. Rev. E}\ }\textbf {\bibinfo {volume} {89}},\
  \bibinfo {pages} {042203} (\bibinfo {year} {2014})}\BibitemShut {NoStop}%
\bibitem [{\citenamefont {Salerno}\ \emph {et~al.}(2018)\citenamefont
  {Salerno}, \citenamefont {Bolintineanu}, \citenamefont {Grest}, \citenamefont
  {Lechman}, \citenamefont {Plimpton}, \citenamefont {Srivastava},\ and\
  \citenamefont {Silbert}}]{salerno2018}%
  \BibitemOpen
  \bibfield  {author} {\bibinfo {author} {\bibfnamefont {K.~M.}\ \bibnamefont
  {Salerno}}, \bibinfo {author} {\bibfnamefont {D.~S.}\ \bibnamefont
  {Bolintineanu}}, \bibinfo {author} {\bibfnamefont {G.~S.}\ \bibnamefont
  {Grest}}, \bibinfo {author} {\bibfnamefont {J.~B.}\ \bibnamefont {Lechman}},
  \bibinfo {author} {\bibfnamefont {S.~J.}\ \bibnamefont {Plimpton}}, \bibinfo
  {author} {\bibfnamefont {I.}~\bibnamefont {Srivastava}},\ and\ \bibinfo
  {author} {\bibfnamefont {L.~E.}\ \bibnamefont {Silbert}},\ }\bibfield
  {title} {\bibinfo {title} {Effect of shape and friction on the packing and
  flow of granular materials},\ }\href@noop {} {\bibfield  {journal} {\bibinfo
  {journal} {Phys. Rev. E}\ }\textbf {\bibinfo {volume} {98}},\ \bibinfo
  {pages} {050901} (\bibinfo {year} {2018})}\BibitemShut {NoStop}%
\bibitem [{\citenamefont {Brouwers}(2006)}]{brouwers2006}%
  \BibitemOpen
  \bibfield  {author} {\bibinfo {author} {\bibfnamefont {H.~J.}\ \bibnamefont
  {Brouwers}},\ }\bibfield  {title} {\bibinfo {title} {Particle-size
  distribution and packing fraction of geometric random packings},\ }\href@noop
  {} {\bibfield  {journal} {\bibinfo  {journal} {Phys. Rev. E}\ }\textbf
  {\bibinfo {volume} {74}},\ \bibinfo {pages} {031309} (\bibinfo {year}
  {2006})}\BibitemShut {NoStop}%
\bibitem [{\citenamefont {Farr}\ and\ \citenamefont {Groot}(2009)}]{farr2009}%
  \BibitemOpen
  \bibfield  {author} {\bibinfo {author} {\bibfnamefont {R.~S.}\ \bibnamefont
  {Farr}}\ and\ \bibinfo {author} {\bibfnamefont {R.~D.}\ \bibnamefont
  {Groot}},\ }\bibfield  {title} {\bibinfo {title} {Close packing density of
  polydisperse hard spheres},\ }\href@noop {} {\bibfield  {journal} {\bibinfo
  {journal} {J. Chem. Phys.}\ }\textbf {\bibinfo {volume} {131}},\ \bibinfo
  {pages} {244104} (\bibinfo {year} {2009})}\BibitemShut {NoStop}%
\bibitem [{\citenamefont {Hopkins}\ \emph {et~al.}(2013)\citenamefont
  {Hopkins}, \citenamefont {Stillinger},\ and\ \citenamefont
  {Torquato}}]{hopkins2013}%
  \BibitemOpen
  \bibfield  {author} {\bibinfo {author} {\bibfnamefont {A.~B.}\ \bibnamefont
  {Hopkins}}, \bibinfo {author} {\bibfnamefont {F.~H.}\ \bibnamefont
  {Stillinger}},\ and\ \bibinfo {author} {\bibfnamefont {S.}~\bibnamefont
  {Torquato}},\ }\bibfield  {title} {\bibinfo {title} {Disordered strictly
  jammed binary sphere packings attain an anomalously large range of
  densities},\ }\href@noop {} {\bibfield  {journal} {\bibinfo  {journal} {Phys.
  Rev. E}\ }\textbf {\bibinfo {volume} {88}},\ \bibinfo {pages} {022205}
  (\bibinfo {year} {2013})}\BibitemShut {NoStop}%
\bibitem [{\citenamefont {Desmond}\ and\ \citenamefont
  {Weeks}(2014)}]{desmond2014}%
  \BibitemOpen
  \bibfield  {author} {\bibinfo {author} {\bibfnamefont {K.~W.}\ \bibnamefont
  {Desmond}}\ and\ \bibinfo {author} {\bibfnamefont {E.~R.}\ \bibnamefont
  {Weeks}},\ }\bibfield  {title} {\bibinfo {title} {Influence of particle size
  distribution on random close packing of spheres},\ }\href@noop {} {\bibfield
  {journal} {\bibinfo  {journal} {Phys. Rev. E}\ }\textbf {\bibinfo {volume}
  {90}},\ \bibinfo {pages} {022204} (\bibinfo {year} {2014})}\BibitemShut
  {NoStop}%
\bibitem [{\citenamefont {Koeze}\ \emph {et~al.}(2016)\citenamefont {Koeze},
  \citenamefont {Vågberg}, \citenamefont {Tjoa},\ and\ \citenamefont
  {Tighe}}]{koeze2016}%
  \BibitemOpen
  \bibfield  {author} {\bibinfo {author} {\bibfnamefont {D.~J.}\ \bibnamefont
  {Koeze}}, \bibinfo {author} {\bibfnamefont {D.}~\bibnamefont {Vågberg}},
  \bibinfo {author} {\bibfnamefont {B.~B.~T.}\ \bibnamefont {Tjoa}},\ and\
  \bibinfo {author} {\bibfnamefont {B.~P.}\ \bibnamefont {Tighe}},\ }\bibfield
  {title} {\bibinfo {title} {Mapping the jamming transition of bidisperse
  mixtures},\ }\href@noop {} {\bibfield  {journal} {\bibinfo  {journal} {EPL
  (Europhys. Lett.)}\ }\textbf {\bibinfo {volume} {113}},\ \bibinfo {pages}
  {54001} (\bibinfo {year} {2016})}\BibitemShut {NoStop}%
\bibitem [{\citenamefont {Prasad}\ \emph {et~al.}(2017)\citenamefont {Prasad},
  \citenamefont {Santangelo},\ and\ \citenamefont {Grason}}]{prasad2017a}%
  \BibitemOpen
  \bibfield  {author} {\bibinfo {author} {\bibfnamefont {I.}~\bibnamefont
  {Prasad}}, \bibinfo {author} {\bibfnamefont {C.}~\bibnamefont {Santangelo}},\
  and\ \bibinfo {author} {\bibfnamefont {G.}~\bibnamefont {Grason}},\
  }\bibfield  {title} {\bibinfo {title} {Subjamming transition in binary sphere
  mixtures},\ }\href@noop {} {\bibfield  {journal} {\bibinfo  {journal} {Phys.
  Rev. E}\ }\textbf {\bibinfo {volume} {96}},\ \bibinfo {pages} {052905}
  (\bibinfo {year} {2017})}\BibitemShut {NoStop}%
\bibitem [{\citenamefont {Pillitteri}\ \emph {et~al.}(2019)\citenamefont
  {Pillitteri}, \citenamefont {Lumay}, \citenamefont {Opsomer},\ and\
  \citenamefont {Vandewalle}}]{pillitteri2019}%
  \BibitemOpen
  \bibfield  {author} {\bibinfo {author} {\bibfnamefont {S.}~\bibnamefont
  {Pillitteri}}, \bibinfo {author} {\bibfnamefont {G.}~\bibnamefont {Lumay}},
  \bibinfo {author} {\bibfnamefont {E.}~\bibnamefont {Opsomer}},\ and\ \bibinfo
  {author} {\bibfnamefont {N.}~\bibnamefont {Vandewalle}},\ }\bibfield  {title}
  {\bibinfo {title} {From jamming to fast compaction dynamics in granular
  binary mixtures},\ }\href@noop {} {\bibfield  {journal} {\bibinfo  {journal}
  {Sci. Rep.}\ }\textbf {\bibinfo {volume} {9}},\ \bibinfo {pages} {1}
  (\bibinfo {year} {2019})}\BibitemShut {NoStop}%
\bibitem [{\citenamefont {Petit}\ \emph {et~al.}(2020)\citenamefont {Petit},
  \citenamefont {Kumar}, \citenamefont {Luding},\ and\ \citenamefont
  {Sperl}}]{petit2020a}%
  \BibitemOpen
  \bibfield  {author} {\bibinfo {author} {\bibfnamefont {J.~C.}\ \bibnamefont
  {Petit}}, \bibinfo {author} {\bibfnamefont {N.}~\bibnamefont {Kumar}},
  \bibinfo {author} {\bibfnamefont {S.}~\bibnamefont {Luding}},\ and\ \bibinfo
  {author} {\bibfnamefont {M.}~\bibnamefont {Sperl}},\ }\bibfield  {title}
  {\bibinfo {title} {Additional transition line in jammed asymmetric bidisperse
  granular packings},\ }\href@noop {} {\bibfield  {journal} {\bibinfo
  {journal} {Phys. Rev. Lett.}\ }\textbf {\bibinfo {volume} {125}},\ \bibinfo
  {pages} {215501} (\bibinfo {year} {2020})}\BibitemShut {NoStop}%
\bibitem [{\citenamefont {Hara}\ \emph {et~al.}(2021)\citenamefont {Hara},
  \citenamefont {Mizuno},\ and\ \citenamefont {Ikeda}}]{hara2020}%
  \BibitemOpen
  \bibfield  {author} {\bibinfo {author} {\bibfnamefont {Y.}~\bibnamefont
  {Hara}}, \bibinfo {author} {\bibfnamefont {H.}~\bibnamefont {Mizuno}},\ and\
  \bibinfo {author} {\bibfnamefont {A.}~\bibnamefont {Ikeda}},\ }\bibfield
  {title} {\bibinfo {title} {Phase transition in the binary mixture of jammed
  particles with large size dispersity},\ }\href@noop {} {\bibfield  {journal}
  {\bibinfo  {journal} {Phys. Rev. Res.}\ }\textbf {\bibinfo {volume} {3}},\
  \bibinfo {pages} {023091} (\bibinfo {year} {2021})}\BibitemShut {NoStop}%
\bibitem [{\citenamefont {Silbert}(2010)}]{silbert2010}%
  \BibitemOpen
  \bibfield  {author} {\bibinfo {author} {\bibfnamefont {L.~E.}\ \bibnamefont
  {Silbert}},\ }\bibfield  {title} {\bibinfo {title} {Jamming of frictional
  spheres and random loose packing},\ }\href@noop {} {\bibfield  {journal}
  {\bibinfo  {journal} {Soft Matter}\ }\textbf {\bibinfo {volume} {6}},\
  \bibinfo {pages} {2918} (\bibinfo {year} {2010})}\BibitemShut {NoStop}%
\bibitem [{\citenamefont {Santos}\ \emph {et~al.}(2020)\citenamefont {Santos},
  \citenamefont {Bolintineanu}, \citenamefont {Grest}, \citenamefont {Lechman},
  \citenamefont {Plimpton}, \citenamefont {Srivastava},\ and\ \citenamefont
  {Silbert}}]{santos2020}%
  \BibitemOpen
  \bibfield  {author} {\bibinfo {author} {\bibfnamefont {A.~P.}\ \bibnamefont
  {Santos}}, \bibinfo {author} {\bibfnamefont {D.~S.}\ \bibnamefont
  {Bolintineanu}}, \bibinfo {author} {\bibfnamefont {G.~S.}\ \bibnamefont
  {Grest}}, \bibinfo {author} {\bibfnamefont {J.~B.}\ \bibnamefont {Lechman}},
  \bibinfo {author} {\bibfnamefont {S.~J.}\ \bibnamefont {Plimpton}}, \bibinfo
  {author} {\bibfnamefont {I.}~\bibnamefont {Srivastava}},\ and\ \bibinfo
  {author} {\bibfnamefont {L.~E.}\ \bibnamefont {Silbert}},\ }\bibfield
  {title} {\bibinfo {title} {Granular packings with sliding, rolling, and
  twisting friction},\ }\href@noop {} {\bibfield  {journal} {\bibinfo
  {journal} {Phys. Rev. E}\ }\textbf {\bibinfo {volume} {102}},\ \bibinfo
  {pages} {032903} (\bibinfo {year} {2020})}\BibitemShut {NoStop}%
\bibitem [{\citenamefont {O’Hern}\ \emph {et~al.}(2003)\citenamefont
  {O’Hern}, \citenamefont {Silbert}, \citenamefont {Liu},\ and\ \citenamefont
  {Nagel}}]{ohern2003}%
  \BibitemOpen
  \bibfield  {author} {\bibinfo {author} {\bibfnamefont {C.~S.}\ \bibnamefont
  {O’Hern}}, \bibinfo {author} {\bibfnamefont {L.~E.}\ \bibnamefont
  {Silbert}}, \bibinfo {author} {\bibfnamefont {A.~J.}\ \bibnamefont {Liu}},\
  and\ \bibinfo {author} {\bibfnamefont {S.~R.}\ \bibnamefont {Nagel}},\
  }\bibfield  {title} {\bibinfo {title} {Jamming at zero temperature and zero
  applied stress: The epitome of disorder},\ }\href@noop {} {\bibfield
  {journal} {\bibinfo  {journal} {Phys. Rev. E}\ }\textbf {\bibinfo {volume}
  {68}},\ \bibinfo {pages} {011306} (\bibinfo {year} {2003})}\BibitemShut
  {NoStop}%
\bibitem [{\citenamefont {Jerkins}\ \emph {et~al.}(2008)\citenamefont
  {Jerkins}, \citenamefont {Schröter}, \citenamefont {Swinney}, \citenamefont
  {Senden}, \citenamefont {Saadatfar},\ and\ \citenamefont
  {Aste}}]{jerkins2008}%
  \BibitemOpen
  \bibfield  {author} {\bibinfo {author} {\bibfnamefont {M.}~\bibnamefont
  {Jerkins}}, \bibinfo {author} {\bibfnamefont {M.}~\bibnamefont {Schröter}},
  \bibinfo {author} {\bibfnamefont {H.~L.}\ \bibnamefont {Swinney}}, \bibinfo
  {author} {\bibfnamefont {T.~J.}\ \bibnamefont {Senden}}, \bibinfo {author}
  {\bibfnamefont {M.}~\bibnamefont {Saadatfar}},\ and\ \bibinfo {author}
  {\bibfnamefont {T.}~\bibnamefont {Aste}},\ }\bibfield  {title} {\bibinfo
  {title} {Onset of mechanical stability in random packings of frictional
  spheres},\ }\href@noop {} {\bibfield  {journal} {\bibinfo  {journal} {Phys.
  Rev. Lett.}\ }\textbf {\bibinfo {volume} {101}},\ \bibinfo {pages} {018301}
  (\bibinfo {year} {2008})}\BibitemShut {NoStop}%
\bibitem [{\citenamefont {Baule}\ \emph {et~al.}(2018)\citenamefont {Baule},
  \citenamefont {Morone}, \citenamefont {Herrmann},\ and\ \citenamefont
  {Makse}}]{baule2018}%
  \BibitemOpen
  \bibfield  {author} {\bibinfo {author} {\bibfnamefont {A.}~\bibnamefont
  {Baule}}, \bibinfo {author} {\bibfnamefont {F.}~\bibnamefont {Morone}},
  \bibinfo {author} {\bibfnamefont {H.~J.}\ \bibnamefont {Herrmann}},\ and\
  \bibinfo {author} {\bibfnamefont {H.~A.}\ \bibnamefont {Makse}},\ }\bibfield
  {title} {\bibinfo {title} {Edwards statistical mechanics for jammed granular
  matter},\ }\href@noop {} {\bibfield  {journal} {\bibinfo  {journal} {Rev.
  Mod. Phys.}\ }\textbf {\bibinfo {volume} {90}},\ \bibinfo {pages} {015006}
  (\bibinfo {year} {2018})}\BibitemShut {NoStop}%
\bibitem [{\citenamefont {Kansal}\ \emph {et~al.}(2002)\citenamefont {Kansal},
  \citenamefont {Torquato},\ and\ \citenamefont {Stillinger}}]{kansal2002}%
  \BibitemOpen
  \bibfield  {author} {\bibinfo {author} {\bibfnamefont {A.~R.}\ \bibnamefont
  {Kansal}}, \bibinfo {author} {\bibfnamefont {S.}~\bibnamefont {Torquato}},\
  and\ \bibinfo {author} {\bibfnamefont {F.~H.}\ \bibnamefont {Stillinger}},\
  }\bibfield  {title} {\bibinfo {title} {Computer generation of dense
  polydisperse sphere packings},\ }\href@noop {} {\bibfield  {journal}
  {\bibinfo  {journal} {J. Chem. Phys.}\ }\textbf {\bibinfo {volume} {117}},\
  \bibinfo {pages} {8212} (\bibinfo {year} {2002})}\BibitemShut {NoStop}%
\bibitem [{\citenamefont {G\"onc\"u}\ and\ \citenamefont
  {Luding}(2013)}]{goncu2013}%
  \BibitemOpen
  \bibfield  {author} {\bibinfo {author} {\bibfnamefont {F.}~\bibnamefont
  {G\"onc\"u}}\ and\ \bibinfo {author} {\bibfnamefont {S.}~\bibnamefont
  {Luding}},\ }\bibfield  {title} {\bibinfo {title} {Effect of particle
  friction and polydispersity on the macroscopic stress–strain relations of
  granular materials},\ }\href {https://doi.org/10.1007/s11440-013-0258-z}
  {\bibfield  {journal} {\bibinfo  {journal} {Acta Geotech.}\ }\textbf
  {\bibinfo {volume} {8}},\ \bibinfo {pages} {629} (\bibinfo {year}
  {2013})}\BibitemShut {NoStop}%
\bibitem [{\citenamefont {Blumenfeld}\ \emph {et~al.}(2005)\citenamefont
  {Blumenfeld}, \citenamefont {Edwards},\ and\ \citenamefont
  {Ball}}]{blumenfeld2005}%
  \BibitemOpen
  \bibfield  {author} {\bibinfo {author} {\bibfnamefont {R.}~\bibnamefont
  {Blumenfeld}}, \bibinfo {author} {\bibfnamefont {S.~F.}\ \bibnamefont
  {Edwards}},\ and\ \bibinfo {author} {\bibfnamefont {R.~C.}\ \bibnamefont
  {Ball}},\ }\bibfield  {title} {\bibinfo {title} {Granular matter and the
  marginal rigidity state},\ }\href@noop {} {\bibfield  {journal} {\bibinfo
  {journal} {J. Condens. Matter Phys.}\ }\textbf {\bibinfo {volume} {17}},\
  \bibinfo {pages} {S2481} (\bibinfo {year} {2005})}\BibitemShut {NoStop}%
\bibitem [{\citenamefont {in~’t Veld}\ \emph {et~al.}(2008)\citenamefont
  {In~’t Veld}, \citenamefont {Plimpton},\ and\ \citenamefont
  {Grest}}]{intveld2008}%
  \BibitemOpen
  \bibfield  {author} {\bibinfo {author} {\bibfnamefont {P.~J.}\ \bibnamefont
  {In~’t Veld}}, \bibinfo {author} {\bibfnamefont {S.~J.}\ \bibnamefont
  {Plimpton}},\ and\ \bibinfo {author} {\bibfnamefont {G.~S.}\ \bibnamefont
  {Grest}},\ }\bibfield  {title} {\bibinfo {title} {Accurate and efficient
  methods for modeling colloidal mixtures in an explicit solvent using
  molecular dynamics},\ }\href@noop {} {\bibfield  {journal} {\bibinfo
  {journal} {Comput. Phys. Commun.}\ }\textbf {\bibinfo {volume} {179}},\
  \bibinfo {pages} {320} (\bibinfo {year} {2008})}\BibitemShut {NoStop}%
\bibitem [{\citenamefont {Ogarko}\ and\ \citenamefont
  {Luding}(2012)}]{ogarko2012}%
  \BibitemOpen
  \bibfield  {author} {\bibinfo {author} {\bibfnamefont {V.}~\bibnamefont
  {Ogarko}}\ and\ \bibinfo {author} {\bibfnamefont {S.}~\bibnamefont
  {Luding}},\ }\bibfield  {title} {\bibinfo {title} {A fast multilevel
  algorithm for contact detection of arbitrarily polydisperse objects},\
  }\href@noop {} {\bibfield  {journal} {\bibinfo  {journal} {Comput. Phys.
  Commun.}\ }\textbf {\bibinfo {volume} {183}},\ \bibinfo {pages} {931–936}
  (\bibinfo {year} {2012})}\BibitemShut {NoStop}%
\bibitem [{\citenamefont {Danisch}\ \emph {et~al.}(2010)\citenamefont
  {Danisch}, \citenamefont {Jin},\ and\ \citenamefont {Makse}}]{danisch2010}%
  \BibitemOpen
  \bibfield  {author} {\bibinfo {author} {\bibfnamefont {M.}~\bibnamefont
  {Danisch}}, \bibinfo {author} {\bibfnamefont {Y.}~\bibnamefont {Jin}},\ and\
  \bibinfo {author} {\bibfnamefont {H.~A.}\ \bibnamefont {Makse}},\ }\bibfield
  {title} {\bibinfo {title} {Model of random packings of different size
  balls},\ }\href@noop {} {\bibfield  {journal} {\bibinfo  {journal} {Phys.
  Rev. E}\ }\textbf {\bibinfo {volume} {81}},\ \bibinfo {pages} {051303}
  (\bibinfo {year} {2010})}\BibitemShut {NoStop}%
\bibitem [{\citenamefont {Cantor}\ \emph {et~al.}(2018)\citenamefont {Cantor},
  \citenamefont {Azéma}, \citenamefont {Sornay},\ and\ \citenamefont
  {Radjai}}]{cantor2018}%
  \BibitemOpen
  \bibfield  {author} {\bibinfo {author} {\bibfnamefont {D.}~\bibnamefont
  {Cantor}}, \bibinfo {author} {\bibfnamefont {E.}~\bibnamefont {Azéma}},
  \bibinfo {author} {\bibfnamefont {P.}~\bibnamefont {Sornay}},\ and\ \bibinfo
  {author} {\bibfnamefont {F.}~\bibnamefont {Radjai}},\ }\bibfield  {title}
  {\bibinfo {title} {Rheology and structure of polydisperse three-dimensional
  packings of spheres},\ }\href@noop {} {\bibfield  {journal} {\bibinfo
  {journal} {Phys. Rev. E}\ }\textbf {\bibinfo {volume} {98}},\ \bibinfo
  {pages} {052910} (\bibinfo {year} {2018})}\BibitemShut {NoStop}%
\bibitem [{\citenamefont {Mutabaruka}\ \emph {et~al.}(2019)\citenamefont
  {Mutabaruka}, \citenamefont {Taiebat}, \citenamefont {Pellenq},\ and\
  \citenamefont {Radjai}}]{mutabaruka2019}%
  \BibitemOpen
  \bibfield  {author} {\bibinfo {author} {\bibfnamefont {P.}~\bibnamefont
  {Mutabaruka}}, \bibinfo {author} {\bibfnamefont {M.}~\bibnamefont {Taiebat}},
  \bibinfo {author} {\bibfnamefont {R.~J.}\ \bibnamefont {Pellenq}},\ and\
  \bibinfo {author} {\bibfnamefont {F.}~\bibnamefont {Radjai}},\ }\bibfield
  {title} {\bibinfo {title} {Effects of size polydispersity on random
  close-packed configurations of spherical particles},\ }\href@noop {}
  {\bibfield  {journal} {\bibinfo  {journal} {Phys. Rev. E}\ }\textbf {\bibinfo
  {volume} {100}},\ \bibinfo {pages} {042906} (\bibinfo {year}
  {2019})}\BibitemShut {NoStop}%
\bibitem [{\citenamefont {Hermes}\ and\ \citenamefont
  {Dijkstra}(2010)}]{hermes2010}%
  \BibitemOpen
  \bibfield  {author} {\bibinfo {author} {\bibfnamefont {M.}~\bibnamefont
  {Hermes}}\ and\ \bibinfo {author} {\bibfnamefont {M.}~\bibnamefont
  {Dijkstra}},\ }\bibfield  {title} {\bibinfo {title} {Jamming of polydisperse
  hard spheres: The effect of kinetic arrest},\ }\href@noop {} {\bibfield
  {journal} {\bibinfo  {journal} {Europhys. Lett.}\ }\textbf {\bibinfo {volume}
  {89}},\ \bibinfo {pages} {38005} (\bibinfo {year} {2010})}\BibitemShut
  {NoStop}%
\bibitem [{\citenamefont {Voigtmann}(2011)}]{voigtmann2011}%
  \BibitemOpen
  \bibfield  {author} {\bibinfo {author} {\bibfnamefont {T.}~\bibnamefont
  {Voigtmann}},\ }\bibfield  {title} {\bibinfo {title} {Multiple glasses in
  asymmetric binary hard spheres},\ }\href@noop {} {\bibfield  {journal}
  {\bibinfo  {journal} {Europhys. Lett.}\ }\textbf {\bibinfo {volume} {96}},\
  \bibinfo {pages} {36006} (\bibinfo {year} {2011})}\BibitemShut {NoStop}%
\bibitem [{\citenamefont {Ricouvier}\ \emph {et~al.}(2017)\citenamefont
  {Ricouvier}, \citenamefont {Pierrat}, \citenamefont {Carminati},
  \citenamefont {Tabeling},\ and\ \citenamefont {Yazhgur}}]{ricouvier2017}%
  \BibitemOpen
  \bibfield  {author} {\bibinfo {author} {\bibfnamefont {J.}~\bibnamefont
  {Ricouvier}}, \bibinfo {author} {\bibfnamefont {R.}~\bibnamefont {Pierrat}},
  \bibinfo {author} {\bibfnamefont {R.}~\bibnamefont {Carminati}}, \bibinfo
  {author} {\bibfnamefont {P.}~\bibnamefont {Tabeling}},\ and\ \bibinfo
  {author} {\bibfnamefont {P.}~\bibnamefont {Yazhgur}},\ }\bibfield  {title}
  {\bibinfo {title} {Optimizing hyperuniformity in self-assembled bidisperse
  emulsions},\ }\href@noop {} {\bibfield  {journal} {\bibinfo  {journal} {Phys.
  Rev. Lett.}\ }\textbf {\bibinfo {volume} {119}},\ \bibinfo {pages} {208001}
  (\bibinfo {year} {2017})}\BibitemShut {NoStop}%
\bibitem [{\citenamefont {Jerolmack}\ and\ \citenamefont
  {Daniels}(2019)}]{jerolmack2019a}%
  \BibitemOpen
  \bibfield  {author} {\bibinfo {author} {\bibfnamefont {D.~J.}\ \bibnamefont
  {Jerolmack}}\ and\ \bibinfo {author} {\bibfnamefont {K.~E.}\ \bibnamefont
  {Daniels}},\ }\bibfield  {title} {\bibinfo {title} {Viewing earth’s surface
  as a soft-matter landscape},\ }\href@noop {} {\bibfield  {journal} {\bibinfo
  {journal} {Nat. Rev. Phys.}\ }\textbf {\bibinfo {volume} {1}},\ \bibinfo
  {pages} {716–730} (\bibinfo {year} {2019})}\BibitemShut {NoStop}%
\bibitem [{\citenamefont {Srivastava}\ \emph {et~al.}(2020)\citenamefont
  {Srivastava}, \citenamefont {Bolintineanu}, \citenamefont {Lechman},\ and\
  \citenamefont {Roberts}}]{srivastava2020a}%
  \BibitemOpen
  \bibfield  {author} {\bibinfo {author} {\bibfnamefont {I.}~\bibnamefont
  {Srivastava}}, \bibinfo {author} {\bibfnamefont {D.~S.}\ \bibnamefont
  {Bolintineanu}}, \bibinfo {author} {\bibfnamefont {J.~B.}\ \bibnamefont
  {Lechman}},\ and\ \bibinfo {author} {\bibfnamefont {S.~A.}\ \bibnamefont
  {Roberts}},\ }\bibfield  {title} {\bibinfo {title} {Controlling binder
  adhesion to impact electrode mesostructures and transport},\ }\href@noop {}
  {\bibfield  {journal} {\bibinfo  {journal} {ACS Appl. Mater. Interfaces}\
  }\textbf {\bibinfo {volume} {12}},\ \bibinfo {pages} {34919} (\bibinfo {year}
  {2020})}\BibitemShut {NoStop}%
\bibitem [{\citenamefont {Masoero}\ \emph {et~al.}(2012)\citenamefont
  {Masoero}, \citenamefont {Del~Gado}, \citenamefont {Pellenq}, \citenamefont
  {Ulm},\ and\ \citenamefont {Yip}}]{masoero2012}%
  \BibitemOpen
  \bibfield  {author} {\bibinfo {author} {\bibfnamefont {E.}~\bibnamefont
  {Masoero}}, \bibinfo {author} {\bibfnamefont {E.}~\bibnamefont {Del~Gado}},
  \bibinfo {author} {\bibfnamefont {R.~J.-M.}\ \bibnamefont {Pellenq}},
  \bibinfo {author} {\bibfnamefont {F.-J.}\ \bibnamefont {Ulm}},\ and\ \bibinfo
  {author} {\bibfnamefont {S.}~\bibnamefont {Yip}},\ }\bibfield  {title}
  {\bibinfo {title} {Nanostructure and nanomechanics of cement: Polydisperse
  colloidal packing},\ }\href@noop {} {\bibfield  {journal} {\bibinfo
  {journal} {Phys. Rev. Lett.}\ }\textbf {\bibinfo {volume} {109}},\ \bibinfo
  {pages} {155503} (\bibinfo {year} {2012})}\BibitemShut {NoStop}%
\bibitem [{\citenamefont {Blanco}\ \emph {et~al.}(2019)\citenamefont {Blanco},
  \citenamefont {Hodgson}, \citenamefont {Hermes}, \citenamefont {Besseling},
  \citenamefont {Hunter}, \citenamefont {Chaikin}, \citenamefont {Cates},
  \citenamefont {Van~Damme},\ and\ \citenamefont {Poon}}]{blanco2019}%
  \BibitemOpen
  \bibfield  {author} {\bibinfo {author} {\bibfnamefont {E.}~\bibnamefont
  {Blanco}}, \bibinfo {author} {\bibfnamefont {D.~J.~M.}\ \bibnamefont
  {Hodgson}}, \bibinfo {author} {\bibfnamefont {M.}~\bibnamefont {Hermes}},
  \bibinfo {author} {\bibfnamefont {R.}~\bibnamefont {Besseling}}, \bibinfo
  {author} {\bibfnamefont {G.~L.}\ \bibnamefont {Hunter}}, \bibinfo {author}
  {\bibfnamefont {P.~M.}\ \bibnamefont {Chaikin}}, \bibinfo {author}
  {\bibfnamefont {M.~E.}\ \bibnamefont {Cates}}, \bibinfo {author}
  {\bibfnamefont {I.}~\bibnamefont {Van~Damme}},\ and\ \bibinfo {author}
  {\bibfnamefont {W.~C.~K.}\ \bibnamefont {Poon}},\ }\bibfield  {title}
  {\bibinfo {title} {Conching chocolate is a prototypical transition from
  frictionally jammed solid to flowable suspension with maximal solid
  content},\ }\href@noop {} {\bibfield  {journal} {\bibinfo  {journal} {Proc.
  Natl. Acad. Sci. U.S.A.}\ }\textbf {\bibinfo {volume} {116}},\ \bibinfo
  {pages} {10303} (\bibinfo {year} {2019})}\BibitemShut {NoStop}%
\bibitem [{\citenamefont {Kumar}\ \emph {et~al.}(2016)\citenamefont {Kumar},
  \citenamefont {Magnanimo}, \citenamefont {Ramaioli},\ and\ \citenamefont
  {Luding}}]{kumar2016a}%
  \BibitemOpen
  \bibfield  {author} {\bibinfo {author} {\bibfnamefont {N.}~\bibnamefont
  {Kumar}}, \bibinfo {author} {\bibfnamefont {V.}~\bibnamefont {Magnanimo}},
  \bibinfo {author} {\bibfnamefont {M.}~\bibnamefont {Ramaioli}},\ and\
  \bibinfo {author} {\bibfnamefont {S.}~\bibnamefont {Luding}},\ }\bibfield
  {title} {\bibinfo {title} {Tuning the bulk properties of bidisperse granular
  mixtures by small amount of fines},\ }\href@noop {} {\bibfield  {journal}
  {\bibinfo  {journal} {Powder Technol.}\ }\textbf {\bibinfo {volume} {293}},\
  \bibinfo {pages} {94} (\bibinfo {year} {2016})}\BibitemShut {NoStop}%
\bibitem [{\citenamefont {Liu}\ \emph {et~al.}(2019)\citenamefont {Liu},
  \citenamefont {Dong}, \citenamefont {Tang}, \citenamefont {Krishnan},
  \citenamefont {Sant},\ and\ \citenamefont {Bauchy}}]{liu2019b}%
  \BibitemOpen
  \bibfield  {author} {\bibinfo {author} {\bibfnamefont {H.}~\bibnamefont
  {Liu}}, \bibinfo {author} {\bibfnamefont {S.}~\bibnamefont {Dong}}, \bibinfo
  {author} {\bibfnamefont {L.}~\bibnamefont {Tang}}, \bibinfo {author}
  {\bibfnamefont {N.~M.~A.}\ \bibnamefont {Krishnan}}, \bibinfo {author}
  {\bibfnamefont {G.}~\bibnamefont {Sant}},\ and\ \bibinfo {author}
  {\bibfnamefont {M.}~\bibnamefont {Bauchy}},\ }\bibfield  {title} {\bibinfo
  {title} {Effects of polydispersity and disorder on the mechanical properties
  of hydrated silicate gels},\ }\href@noop {} {\bibfield  {journal} {\bibinfo
  {journal} {J. Mech. Phys. Solids}\ }\textbf {\bibinfo {volume} {122}},\
  \bibinfo {pages} {555} (\bibinfo {year} {2019})}\BibitemShut {NoStop}%
\bibitem [{\citenamefont {Spangenberg}\ \emph {et~al.}(2014)\citenamefont
  {Spangenberg}, \citenamefont {Scherer}, \citenamefont {Hopkins},\ and\
  \citenamefont {Torquato}}]{spangenberg2014}%
  \BibitemOpen
  \bibfield  {author} {\bibinfo {author} {\bibfnamefont {J.}~\bibnamefont
  {Spangenberg}}, \bibinfo {author} {\bibfnamefont {G.~W.}\ \bibnamefont
  {Scherer}}, \bibinfo {author} {\bibfnamefont {A.~B.}\ \bibnamefont
  {Hopkins}},\ and\ \bibinfo {author} {\bibfnamefont {S.}~\bibnamefont
  {Torquato}},\ }\bibfield  {title} {\bibinfo {title} {Viscosity of bimodal
  suspensions with hard spherical particles},\ }\href@noop {} {\bibfield
  {journal} {\bibinfo  {journal} {J. Appl. Phys.}\ }\textbf {\bibinfo {volume}
  {116}},\ \bibinfo {pages} {184902} (\bibinfo {year} {2014})}\BibitemShut
  {NoStop}%
\bibitem [{\citenamefont {Pednekar}\ \emph {et~al.}(2018)\citenamefont
  {Pednekar}, \citenamefont {Chun},\ and\ \citenamefont
  {Morris}}]{pednekar2018}%
  \BibitemOpen
  \bibfield  {author} {\bibinfo {author} {\bibfnamefont {S.}~\bibnamefont
  {Pednekar}}, \bibinfo {author} {\bibfnamefont {J.}~\bibnamefont {Chun}},\
  and\ \bibinfo {author} {\bibfnamefont {J.~F.}\ \bibnamefont {Morris}},\
  }\bibfield  {title} {\bibinfo {title} {Bidisperse and polydisperse suspension
  rheology at large solid fraction},\ }\href@noop {} {\bibfield  {journal}
  {\bibinfo  {journal} {J. Rheol.}\ }\textbf {\bibinfo {volume} {62}},\
  \bibinfo {pages} {513} (\bibinfo {year} {2018})}\BibitemShut {NoStop}%
\bibitem [{\citenamefont {Guy}\ \emph {et~al.}(2020)\citenamefont {Guy},
  \citenamefont {Ness}, \citenamefont {Hermes}, \citenamefont {Sawiak},
  \citenamefont {Sun},\ and\ \citenamefont {Poon}}]{guy2020}%
  \BibitemOpen
  \bibfield  {author} {\bibinfo {author} {\bibfnamefont {B.~M.}\ \bibnamefont
  {Guy}}, \bibinfo {author} {\bibfnamefont {C.}~\bibnamefont {Ness}}, \bibinfo
  {author} {\bibfnamefont {M.}~\bibnamefont {Hermes}}, \bibinfo {author}
  {\bibfnamefont {L.~J.}\ \bibnamefont {Sawiak}}, \bibinfo {author}
  {\bibfnamefont {J.}~\bibnamefont {Sun}},\ and\ \bibinfo {author}
  {\bibfnamefont {W.~C.~K.}\ \bibnamefont {Poon}},\ }\bibfield  {title}
  {\bibinfo {title} {Testing the wyart-cates model for non-brownian shear
  thickening using bidisperse suspensions.},\ }\href@noop {} {\bibfield
  {journal} {\bibinfo  {journal} {Soft Matter}\ }\textbf {\bibinfo {volume}
  {16}},\ \bibinfo {pages} {229} (\bibinfo {year} {2020})}\BibitemShut
  {NoStop}%
\bibitem [{\citenamefont {Furnas}(1931)}]{furnas1929}%
  \BibitemOpen
  \bibfield  {author} {\bibinfo {author} {\bibfnamefont {C.~C.}\ \bibnamefont
  {Furnas}},\ }\bibfield  {title} {\bibinfo {title} {Grading aggregates i.
  mathematical relations for beds of broken solids of maximum density},\
  }\href@noop {} {\bibfield  {journal} {\bibinfo  {journal} {Ind. Eng. Chem.
  Res.}\ }\textbf {\bibinfo {volume} {23}},\ \bibinfo {pages} {1052} (\bibinfo
  {year} {1931})}\BibitemShut {NoStop}%
\bibitem [{\citenamefont {Dagois-Bohy}\ \emph {et~al.}(2012)\citenamefont
  {Dagois-Bohy}, \citenamefont {Tighe}, \citenamefont {Simon}, \citenamefont
  {Henkes},\ and\ \citenamefont {Van~Hecke}}]{dagois-bohy2012}%
  \BibitemOpen
  \bibfield  {author} {\bibinfo {author} {\bibfnamefont {S.}~\bibnamefont
  {Dagois-Bohy}}, \bibinfo {author} {\bibfnamefont {B.~P.}\ \bibnamefont
  {Tighe}}, \bibinfo {author} {\bibfnamefont {J.}~\bibnamefont {Simon}},
  \bibinfo {author} {\bibfnamefont {S.}~\bibnamefont {Henkes}},\ and\ \bibinfo
  {author} {\bibfnamefont {M.}~\bibnamefont {Van~Hecke}},\ }\bibfield  {title}
  {\bibinfo {title} {Soft-sphere packings at finite pressure but unstable to
  shear},\ }\href@noop {} {\bibfield  {journal} {\bibinfo  {journal} {Phys.
  Rev. Lett.}\ }\textbf {\bibinfo {volume} {109}},\ \bibinfo {pages} {095703}
  (\bibinfo {year} {2012})}\BibitemShut {NoStop}%
\bibitem [{\citenamefont {Plimpton}(1995)}]{plimpton1995}%
  \BibitemOpen
  \bibfield  {author} {\bibinfo {author} {\bibfnamefont {S.}~\bibnamefont
  {Plimpton}},\ }\bibfield  {title} {\bibinfo {title} {Fast parallel algorithms
  for short-range molecular dynamics},\ }\href@noop {} {\bibfield  {journal}
  {\bibinfo  {journal} {J. Comput. Phys.}\ }\textbf {\bibinfo {volume} {117}},\
  \bibinfo {pages} {1} (\bibinfo {year} {1995})}\BibitemShut {NoStop}%
\bibitem [{\citenamefont {Srivastava}\ \emph {et~al.}(2021)\citenamefont
  {Srivastava}, \citenamefont {Silbert}, \citenamefont {Grest},\ and\
  \citenamefont {Lechman}}]{srivastava2021}%
  \BibitemOpen
  \bibfield  {author} {\bibinfo {author} {\bibfnamefont {I.}~\bibnamefont
  {Srivastava}}, \bibinfo {author} {\bibfnamefont {L.~E.}\ \bibnamefont
  {Silbert}}, \bibinfo {author} {\bibfnamefont {G.~S.}\ \bibnamefont {Grest}},\
  and\ \bibinfo {author} {\bibfnamefont {J.~B.}\ \bibnamefont {Lechman}},\
  }\bibfield  {title} {\bibinfo {title} {Viscometric flow of dense granular
  materials under controlled pressure and shear stress},\ }\href@noop {}
  {\bibfield  {journal} {\bibinfo  {journal} {J. Fluid Mech.}\ }\textbf
  {\bibinfo {volume} {907}},\ \bibinfo {pages} {A18} (\bibinfo {year}
  {2021})}\BibitemShut {NoStop}%
\bibitem [{\citenamefont {Srivastava}\ \emph {et~al.}(2019)\citenamefont
  {Srivastava}, \citenamefont {Silbert}, \citenamefont {Grest},\ and\
  \citenamefont {Lechman}}]{srivastava2019}%
  \BibitemOpen
  \bibfield  {author} {\bibinfo {author} {\bibfnamefont {I.}~\bibnamefont
  {Srivastava}}, \bibinfo {author} {\bibfnamefont {L.~E.}\ \bibnamefont
  {Silbert}}, \bibinfo {author} {\bibfnamefont {G.~S.}\ \bibnamefont {Grest}},\
  and\ \bibinfo {author} {\bibfnamefont {J.~B.}\ \bibnamefont {Lechman}},\
  }\bibfield  {title} {\bibinfo {title} {Flow-arrest transitions in frictional
  granular matter},\ }\href@noop {} {\bibfield  {journal} {\bibinfo  {journal}
  {Phys. Rev. Lett.}\ }\textbf {\bibinfo {volume} {122}},\ \bibinfo {pages}
  {048003} (\bibinfo {year} {2019})}\BibitemShut {NoStop}%
\bibitem [{\citenamefont {Roux}(2000)}]{roux2000}%
  \BibitemOpen
  \bibfield  {author} {\bibinfo {author} {\bibfnamefont {J.-N.}\ \bibnamefont
  {Roux}},\ }\bibfield  {title} {\bibinfo {title} {Geometric origin of
  mechanical properties of granular materials},\ }\href@noop {} {\bibfield
  {journal} {\bibinfo  {journal} {Phys. Rev. E}\ }\textbf {\bibinfo {volume}
  {61}},\ \bibinfo {pages} {6802} (\bibinfo {year} {2000})}\BibitemShut
  {NoStop}%
\bibitem [{\citenamefont {Srivastava}\ and\ \citenamefont
  {Fisher}(2017)}]{srivastava2017}%
  \BibitemOpen
  \bibfield  {author} {\bibinfo {author} {\bibfnamefont {I.}~\bibnamefont
  {Srivastava}}\ and\ \bibinfo {author} {\bibfnamefont {T.~S.}\ \bibnamefont
  {Fisher}},\ }\bibfield  {title} {\bibinfo {title} {Slow creep in soft
  granular packings},\ }\href@noop {} {\bibfield  {journal} {\bibinfo
  {journal} {Soft Matter}\ }\textbf {\bibinfo {volume} {13}},\ \bibinfo {pages}
  {3411} (\bibinfo {year} {2017})}\BibitemShut {NoStop}%
\bibitem [{Note1()}]{Note1}%
  \BibitemOpen
  \bibinfo {note} {We used a recently proposed algorithm to speedup particle
  contact detection~\cite {shire2020}. Separate spatial binning of small and
  large particles reduces the time to find potential neighbors, while a
  hierarchical search from small to large particles reduces the interprocessor
  communication.}\BibitemShut {Stop}%
\bibitem [{\citenamefont {Luding}(2016)}]{luding2016}%
  \BibitemOpen
  \bibfield  {author} {\bibinfo {author} {\bibfnamefont {S.}~\bibnamefont
  {Luding}},\ }\bibfield  {title} {\bibinfo {title} {So much for the jamming
  point},\ }\href@noop {} {\bibfield  {journal} {\bibinfo  {journal} {Nat.
  Phys.}\ }\textbf {\bibinfo {volume} {12}},\ \bibinfo {pages} {531} (\bibinfo
  {year} {2016})}\BibitemShut {NoStop}%
\bibitem [{\citenamefont {Bertrand}\ \emph {et~al.}(2016)\citenamefont
  {Bertrand}, \citenamefont {Behringer}, \citenamefont {Chakraborty},
  \citenamefont {O’Hern},\ and\ \citenamefont {Shattuck}}]{bertrand2016}%
  \BibitemOpen
  \bibfield  {author} {\bibinfo {author} {\bibfnamefont {T.}~\bibnamefont
  {Bertrand}}, \bibinfo {author} {\bibfnamefont {R.~P.}\ \bibnamefont
  {Behringer}}, \bibinfo {author} {\bibfnamefont {B.}~\bibnamefont
  {Chakraborty}}, \bibinfo {author} {\bibfnamefont {C.~S.}\ \bibnamefont
  {O’Hern}},\ and\ \bibinfo {author} {\bibfnamefont {M.~D.}\ \bibnamefont
  {Shattuck}},\ }\bibfield  {title} {\bibinfo {title} {Protocol dependence of
  the jamming transition},\ }\href@noop {} {\bibfield  {journal} {\bibinfo
  {journal} {Phys. Rev. E}\ }\textbf {\bibinfo {volume} {93}},\ \bibinfo
  {pages} {012901} (\bibinfo {year} {2016})}\BibitemShut {NoStop}%
\bibitem [{\citenamefont {Donev}\ \emph {et~al.}(2004)\citenamefont {Donev},
  \citenamefont {Torquato}, \citenamefont {Stillinger},\ and\ \citenamefont
  {Connelly}}]{donev2004}%
  \BibitemOpen
  \bibfield  {author} {\bibinfo {author} {\bibfnamefont {A.}~\bibnamefont
  {Donev}}, \bibinfo {author} {\bibfnamefont {S.}~\bibnamefont {Torquato}},
  \bibinfo {author} {\bibfnamefont {F.~H.}\ \bibnamefont {Stillinger}},\ and\
  \bibinfo {author} {\bibfnamefont {R.}~\bibnamefont {Connelly}},\ }\bibfield
  {title} {\bibinfo {title} {A linear programming algorithm to test for jamming
  in hard-sphere packings},\ }\href@noop {} {\bibfield  {journal} {\bibinfo
  {journal} {J. Comput. Phys.}\ }\textbf {\bibinfo {volume} {197}},\ \bibinfo
  {pages} {139} (\bibinfo {year} {2004})}\BibitemShut {NoStop}%
\bibitem [{\citenamefont {Shundyak}\ \emph {et~al.}(2007)\citenamefont
  {Shundyak}, \citenamefont {Van~Hecke},\ and\ \citenamefont
  {Van~Saarloos}}]{shundyak2007}%
  \BibitemOpen
  \bibfield  {author} {\bibinfo {author} {\bibfnamefont {K.}~\bibnamefont
  {Shundyak}}, \bibinfo {author} {\bibfnamefont {M.}~\bibnamefont
  {Van~Hecke}},\ and\ \bibinfo {author} {\bibfnamefont {W.}~\bibnamefont
  {Van~Saarloos}},\ }\bibfield  {title} {\bibinfo {title} {Force mobilization
  and generalized isostaticity in jammed packings of frictional grains.},\
  }\href@noop {} {\bibfield  {journal} {\bibinfo  {journal} {Phys. Rev. E}\
  }\textbf {\bibinfo {volume} {75}},\ \bibinfo {pages} {010301} (\bibinfo
  {year} {2007})}\BibitemShut {NoStop}%
\bibitem [{\citenamefont {Henkes}\ \emph {et~al.}(2016)\citenamefont {Henkes},
  \citenamefont {Quint}, \citenamefont {Fily},\ and\ \citenamefont
  {Schwarz}}]{henkes2016}%
  \BibitemOpen
  \bibfield  {author} {\bibinfo {author} {\bibfnamefont {S.}~\bibnamefont
  {Henkes}}, \bibinfo {author} {\bibfnamefont {D.~A.}\ \bibnamefont {Quint}},
  \bibinfo {author} {\bibfnamefont {Y.}~\bibnamefont {Fily}},\ and\ \bibinfo
  {author} {\bibfnamefont {J.~M.}\ \bibnamefont {Schwarz}},\ }\bibfield
  {title} {\bibinfo {title} {Rigid cluster decomposition reveals criticality in
  frictional jamming},\ }\href@noop {} {\bibfield  {journal} {\bibinfo
  {journal} {Phys. Rev. Lett.}\ }\textbf {\bibinfo {volume} {116}},\ \bibinfo
  {pages} {028301} (\bibinfo {year} {2016})}\BibitemShut {NoStop}%
\bibitem [{\citenamefont {Liu}\ \emph {et~al.}(2021)\citenamefont {Liu},
  \citenamefont {Kollmer}, \citenamefont {Daniels}, \citenamefont {Schwarz},\
  and\ \citenamefont {Henkes}}]{liu2021}%
  \BibitemOpen
  \bibfield  {author} {\bibinfo {author} {\bibfnamefont {K.}~\bibnamefont
  {Liu}}, \bibinfo {author} {\bibfnamefont {J.~E.}\ \bibnamefont {Kollmer}},
  \bibinfo {author} {\bibfnamefont {K.~E.}\ \bibnamefont {Daniels}}, \bibinfo
  {author} {\bibfnamefont {J.~M.}\ \bibnamefont {Schwarz}},\ and\ \bibinfo
  {author} {\bibfnamefont {S.}~\bibnamefont {Henkes}},\ }\bibfield  {title}
  {\bibinfo {title} {Spongelike rigid structures in frictional granular
  packings},\ }\href@noop {} {\bibfield  {journal} {\bibinfo  {journal} {Phys.
  Rev. Lett.}\ }\textbf {\bibinfo {volume} {126}},\ \bibinfo {pages} {088002}
  (\bibinfo {year} {2021})}\BibitemShut {NoStop}%
\bibitem [{\citenamefont {Shire}\ \emph {et~al.}(2020)\citenamefont {Shire},
  \citenamefont {Hanley},\ and\ \citenamefont {Stratford}}]{shire2020}%
  \BibitemOpen
  \bibfield  {author} {\bibinfo {author} {\bibfnamefont {T.}~\bibnamefont
  {Shire}}, \bibinfo {author} {\bibfnamefont {K.~J.}\ \bibnamefont {Hanley}},\
  and\ \bibinfo {author} {\bibfnamefont {K.}~\bibnamefont {Stratford}},\
  }\bibfield  {title} {\bibinfo {title} {DEM simulations of polydisperse media:
  efficient contact detection applied to investigate the quasi-static limit},\
  }\href@noop {} {\bibfield  {journal} {\bibinfo  {journal} {Comput. Part.
  Mech.}\ }\textbf {\bibinfo {volume} {8}},\ \bibinfo {pages} {653} (\bibinfo
  {year} {2020})}\BibitemShut {NoStop}%
\end{thebibliography}
\end{document}